\documentstyle[twoside,fleqn,espcrc2,epsfig]{article}

\newcommand{\AmS}{{\protect\the\textfont2
  A\kern-.1667em\lower.5ex\hbox{M}\kern-.125emS}}

\title{Debye mass from domainwalls and dimensionally reduced phase diagram}

\author{S. Bronoff
        and 
        C. P. Korthals Altes
\address{Centre Pysique Th\'eorique au CNRS,\\
Case 907, Luminy, F13288, Marseille, France }}
       
\begin{document}

\begin{abstract}
To measure the Debye mass in dimensionally
reduced QCD for $N_c \ge 3$ we replace in the  correlator of two Polyakov loops 
one of the loops by a wall triggered by a  dimensionally reduced twist. The phase diagram for $N_c=3$ has R-parity broken in part of the Higgs phase.  
\end{abstract}

\maketitle

\section{Introduction}
Precise numerical knowledge of  QCD plasma parameters are of paramount importance in view of the imminence of data from RHIC. The Debye screening mass plays a crucial role in determining typical signals from the eventual plasma formation,  like in J/$\Psi$ suppression. We have, at temperatures a few times higher than the critical one, the possibility of making estimates which are partly analytic, partly numerical through dimensional reduction. One uses perturbationtheory to integrate out the heavy ($\sim T$) degrees of freedom. One is then left with an effective action containg only the light ($\le T$) static  degrees of freedom, living in one  dimension less\cite{dimred}.
Numerically the advantage of having a three instead of four dimensional theory is enormous. In this note we describe a particular observable (a ``wall''), that triggers physical, i.e. gauge invariant excitations and one of them is the 
excitation carrying as lowest mass the  Debye mass. This excitation is subsequently captured by the imaginary part of the Wilson line at sufficiently large distance. So we need at least three colours. A one loop calculation shows
that for small x R-partity\cite{debye} is broken in the Higgs phase.

\section{The observable}

In this section we will describe the observable in question.
We start with the reduced 3D version of QCD: it consistst of a 3D gauge part $S_G$ and a part $S_A$, where the adjoint $A_0$ couples.
\begin{equation}
S_G=\sum_{i,j,\vec x}F_{ij}^2
\end{equation}
and
\begin{eqnarray}
S_A&=&\sum_{i,\vec x}Tr(D_iA_0)^2\nonumber\\
&+&\sum_{\vec x}\left(m_E^2TrA_0^2+2\lambda_A TrA_0^4\right)+\delta S
\end{eqnarray}

The first term is the colour-magnetic term, the second the static colour electric term. The third and fourth term contain the effects of the perturbative
integration of the heavy modes in the gauge fields and  fermions. They have been computed to two loop order~\cite{loop}. The two loop integrations introduce a subtraction scale that one usualy chooses such that the renormalisation of the three dimensional gauge coupling $g_3$ is simplest~\cite{huang}.

$\delta S$ contains the sextic and higher order terms. To one and two loop order these are all zero~\cite{these}. 

The latticized form of the action is given by the standard plaquette action for the magnetic term:
\begin{equation}
S_P=\beta\sum_{ij,\vec x}\left(1-{1\over N}Re Tr P_{ij}(U)(\vec x)\right), 
\end{equation}
and in addition the kinetic and potential terms for the adjoint scalar $A_0$:
\begin{eqnarray}
S_A&=&-2\sum_{i,\vec x}Tr\left(A_0(\vec x)U_i(\vec x)A_0(\vec x+\vec e_i)U^{\dagger}_i(\vec x)\right)\nonumber\\
&+&\sum_{\vec x}\left((6+m_E^2)TrA_0^2+2\lambda_A TrA_0^4\right)
\end{eqnarray}
All quantities are in units of the lattice spacing. $\vec x$ stands for integer valued three vectors.
Our system is enclosed in a box of size $L_x=L_y=L_{tr}\ll L_z$. Our boundary conditions are periodic in all directions. This is important for the physics in the next section.

The continuum limit of this superrenormalizable theory is governed
by one and two-loop divergencies in the mass term. Thus the tree level
relations between the coupling on the lattice and in the continuum
are affected in a straightforward fashion\cite{laine}. We are interested
in the limit $\beta\to \infty$, $\lambda\to 0$, whilst keeping the dimensionless continuum variables $x=\lambda_A/{g_3^2}$ and $y=m_E^2/{g_3^4}$ fixed.

Now our observable. We select an (x-y) plane say at $z={1\over 2}$.
It cuts a planar array of links in the z-direction, all starting at z=0 and ending at z=1. On both ends of these links,{\it{ and only there}}, we are going to change
the adjoint variable $A_0$ into $\exp{iA_0}$. The kinetic term for the exponentiated adjoint is then multiplied with an element $\omega$ of the centergroup
of SU(N), say $\omega=\exp{i2\pi/N}$. We have to add the hermitian conjugate to the terms where the exponentiated field appears. Explicitely, for the links in question: 
\begin{eqnarray}
S_t&=&-2\omega Tr\left(e^{iA_0(\vec o)}U_z(\vec 0)e^{iA_0(001)}U_z(\vec 0)\right)\nonumber\\
&+&\omega\left(Tre^{2iA_0(\vec o)}+Tre^{2iA_0(001)}\right)+h.c.
\end{eqnarray}
This is all the change we make in the action, and the result is called the
twisted action $S_t$. The use of twist\cite{groe,wall4d} has been useful in 4D pure Yang-Mills.
 
\section{Physics of the wall}

What purpose does the twisted action serve? When we are approaching the 
continuum limit as discussed above ($U_i\sim 1$)  in the original action the kinetic term goes to zero. This is certainly true in a phase where the expectation value of $A_0$ is zero. However, the twisted action equals
\begin{equation}
S_t= 2(1-Re\omega)L_{tr}^2
\end{equation}
 That is, the twisted system builds up a wall, when we are approaching the continuum limit as discussed above.
 
\subsection{Excitations of the wall in the vacuum}
Far away from the wall we have the vacuum state, which extends through the periodic boundary conditions also to the other side of the wall.
This is important: the wall falls off into this vacuum, and excites therefore the vacuum near the wall. It is these excitations that contain the Debye mass. Under R-symmetry\cite{debye} the wall contains excitations of the quantum number of the Debye mass:it has an odd component under $A_0\to -A_0$ and $A_i$.

\subsection{Which vacuum do we choose?}

\begin{figure}[htb]
\epsfig{figure=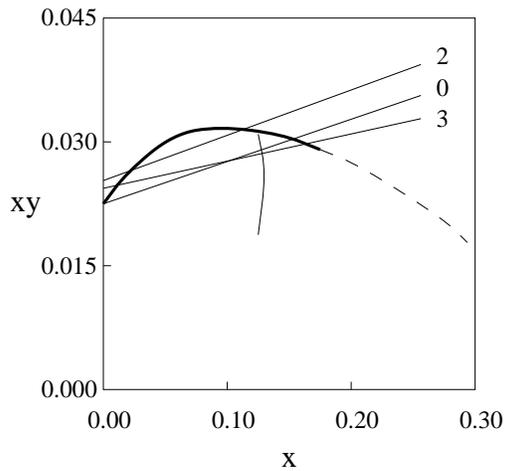,width=6.5cm,angle=0}
\caption{Sketch of phase diagram of 3d theory in continuum limit with lines of 4d physics for various numbers of fermions. The thin continuous curve indicates where the cubic invariant may become zero.}
\label{fig:largenenough}
\end{figure}
The phase diagram in fig. 1 is based on a measurement of the
average of $Tr A_0^2$ with of course the untwisted action S. In absence of such a measurement we sketched the critical line. Above the solid line the average is zero, and we have a symmetric phase. Below we have the Higgs phase where the average is non-zero. For small enough x, and for all $N_c$ a one loop calculation of the effective potential as function of all independent invariants shows the well known $Z(N_c)$ structure~\cite{these,karsch}. For $N_c=2$ this was done in ref.~\cite{loop}. For N=3 or larger we minimize the effective potential keeping either the quadratic or cubic invariant fixed. The minima are in both cases in the
locations where the following equality is valid:
\begin{equation}
TrA_0^3=\pm{(N_c-2)\over\sqrt{N_c(N_c-1)}}\left(TrA_0^2\right)^{3/2}
\end{equation}
So when the quadratic VEV jumps, the cubic one has to follow! Hence we want to approach the physics lines from the {\it{symmetric}} vacuum. Part of the 4D physics lines lie in the metastability range of the transition. So in that case  we can only measure the physics in 4D in that part of the phase diagram, where the nucleation time is much larger
than the Montecarlo thermalization time~\cite{loop}.

\section{Extracting the Debye mass}

Finally we have to give a method to extract the Debye mass. This can be done by computing the expectation value of the imaginary part $Im L$of the trace of the Wilson
line with the twisted action. This operator is odd under $A_0\to -A_0$.
Any other one is equally good.
Denote this average by $\langle Im\bar L(z)\rangle_{S_t}$. The bar means
we have averaged the loop L over the (x-y) plane at position z. Then we have in
an obvious notation the following trivial relation:
\begin{equation}
\langle Im \bar L\rangle_{S_t}=\langle\exp{-(S_t-S)}Im\bar L(z)\rangle_{S}
\end{equation}
In other words the twisted average of the Wilson line is the correlation between the
wall operator $\exp{-(S_t-S)}$ at z=0 and the Wilson line at z.
In order to see the fall-off due to the lowest mass excitation we have to
take z far away from the wall:
\begin{equation}
\langle Im\bar L(z)\rangle_{S_t}\sim\exp{-m_D\vert z\vert}
\end{equation}

\section{Conclusions}
In conclusion, we have shown how to obtain the Debye mass from the fall-off of a domainwall. The domainwall is obtained by imposing a twist in the action, which is easy to implement in a Montecarlo simulation. 

The coupling , or overlap, of the wall to the state carrying the Debye mass is an important issue. 
This has to be established by empirical means. 

The phase diagram for $N_c=3$ as sketched in the figure breaks R-symmetry for
small x. But the relation between the relevant VEV's may breakdown in the Higgsphase at the thin line. The phase diagram should be 
searched numerically for this phenomenon.

Domainwalls play an important role in the deconfining transition. Until now they defy the use of
reduction methods because the Wilson line fluctuates on the order of one.
Here we look only at exponentially small values of the line far in the 
bulk vacuum. But this way of looking at the problem may lead to progress
in 3D understanding of the phase transition.

\end{document}